\documentclass[11pt]{article}

\usepackage[margin=1in]{geometry}
\usepackage{amsmath,amssymb,amsfonts}
\usepackage{graphicx}
\usepackage{authblk,natbib}
\usepackage{setspace}
\usepackage{bbm}
\usepackage{xcolor}
\usepackage[bookmarksopen=true,bookmarks=true,pdfencoding=auto,psdextra,colorlinks]{hyperref}
\hypersetup{ 
    colorlinks,
    linkcolor={blue!80!black},
    citecolor={green!40!black},
    urlcolor={red!50!black}
}

\let\hat\widehat
\let\tilde\widetilde

\title{Effects of Distance Metrics and Scaling on the Perturbation Discrimination Score}

\author[1]{Qiyuan Liu}
\author[2]{Qirui Zhang}
\author[3]{Jinhong Du$^*$}
\author[2]{Siming Zhao$^*$}
\author[1]{Jingshu Wang\thanks{Correspondence: 
\texttt{jingshuw@uchicago.edu};
\texttt{siming.zhao@dartmouth.edu}; 
\texttt{jinhongd@hku.hk}}
}

\affil[1]{Department of Statistics, University of Chicago, Chicago, IL, USA}
\affil[2]{Department of Biomedical Data Science, Dartmouth College, Hanover, NH, USA; Dartmouth Cancer Center, Lebanon, NH, USA}
\affil[3]{Institute of Data Science, The University of Hong Kong, Hong Kong SAR, China; Department of Statistics and Actuarial Science, The University of Hong Kong, Hong Kong SAR, China}

\date{}

\begin{document}

\maketitle

\begin{abstract}
    The Perturbation Discrimination Score (PDS) is increasingly used to evaluate whether predicted perturbation effects remain distinguishable, including in Systema and the Virtual Cell Challenge. However, its behavior in high-dimensional gene-expression settings has not been examined in detail. We show that PDS is highly sensitive to the choice of similarity or distance measure and to the scale of predicted effects. Analysis of observed perturbation responses reveals that $\ell_1$ and $\ell_2$-based PDS behave very differently from cosine-based measures, even after norm matching. We provide geometric insight and discuss implications for future discrimination-based evaluation metrics.
\end{abstract}

\section*{Introduction}

A central goal in functional genomics is to understand how genetic perturbations change gene expression. 
When predicting perturbation effects, an important question is not only how close the predictions are on average, but also whether different perturbations remain distinguishable. The Perturbation Discrimination Score (PDS) formalizes this idea by evaluating whether a predicted perturbation-effect vector $\hat \Delta_i$ is closest to its true counterpart $\Delta_i$ for perturbation $i$ within a collection of perturbations. Metrics of this form have been used in several recent benchmarking efforts. For example, the Systema framework \citep{vinas2025systema} evaluates discriminability using centroid accuracy, which corresponds to PDS defined with the Euclidean ($\ell_2$) distance. The PerturbBench framework \citep{wu2024perturbench} develops a class of rank-based metrics including PDS under generic distance definitions, and the Virtual Cell Challenge (VCC; \cite{roohani2025virtual}) further adopts this PDS metric with the Manhattan ($\ell_1$) distance for large-scale benchmarking.

Although discrimination-based metrics are motivated by clear biological considerations and are increasingly used in benchmarking studies, they remain relatively new in prediction problems, and their behavior under different distance or similarity definitions has not been systematically examined.   Understanding these properties is important for interpreting benchmark results and for designing robust metrics for future perturbation-prediction studies. In our analysis, we find that the behavior of PDS depends strongly on the choice of similarity or distance measure. In particular, PDS defined using $\ell_1/\ell_2$ distance is highly sensitive to the scale of predicted effects, whereas cosine-based versions are not, leading to interactions between metric choice and scale that are not always intuitive.

In this commentary, we analyze how PDS behaves under different distance measures, with a focus on how directional agreement, magnitude, and scaling together influence discriminability. Our goal is to clarify how PDS behaves in practice and to provide insight that may guide the development of future discrimination-based evaluation metrics.

\section*{Different distance metrics greatly change PDS performance}

At its core, for each perturbation $i$, PDS ranks the true distance $d(\hat{\Delta}_i, \Delta_{i})$ among the set of distances $d(\hat{\Delta}_i, \Delta_{i'})$ for all perturbation $i'$, and linearly rescales that rank to the interval $[0, 1]$. A perfect match gives PDS $=1$ and random guessing gives roughly $0.5$, with the worst case giving $0$.

Systema used the $\ell_2$ distance: 
\[
d_{\ell_2}(\hat{\Delta}_i,\Delta_{i'}) =  \|\hat{\Delta}_{i} - \Delta_{i'}\|_2 = \sqrt{\sum_j (\hat{\Delta}_{ij} - \Delta_{i'j})^2},
\]
and VCC used the $\ell_1$ distance:
\[
d_{\ell_1}(\hat{\Delta}_i,\Delta_{i'}) = \|\hat{\Delta}_{i} - \Delta_{i'}\|_1 = \sum_j |\hat{\Delta}_{ij} - \Delta_{i'j}|.
\]
However, we find that PDS can vary dramatically depending on the choice of distance.

To illustrate this effect, we consider a simple prediction strategy: using the genome-wide CRISPR screen in the K562 cell line \cite{replogle2022mapping} to predict the mean perturbation effects in the hESC training set provided by the VCC. We restrict the comparison to the 115 perturbations and 7,581 genes shared between the two datasets.
For each perturbation $i$, the predicted $\hat{\Delta}_i$ is defined as the mean difference between perturbed and control cells in the K562 data after standard preprocessing. The corresponding ``true'' effect ${\Delta}_i$ is defined analogously using the VCC hESC training data, ensuring that both sets of effects are computed under the same preprocessing pipeline.

We then compute PDS under four definitions of distances: 
\begin{enumerate}
    \item[\textbf{(1)}] $\ell_1$ distance,
    \item[\textbf{(2)}] $\ell_2$ distance,
    \item[\textbf{(3)}] cosine dissimilarity $1 - \cos(\hat{\Delta}_i,\Delta_{i'})$,
    \item[\textbf{(4)}] sign-based cosine dissimilarity $1 - \cos(\text{sign}(\hat{\Delta}_i),\text{sign}(\Delta_{i'}))$.
\end{enumerate}

\begin{figure}[ht]
    \centering
    \includegraphics[width=0.9\linewidth]{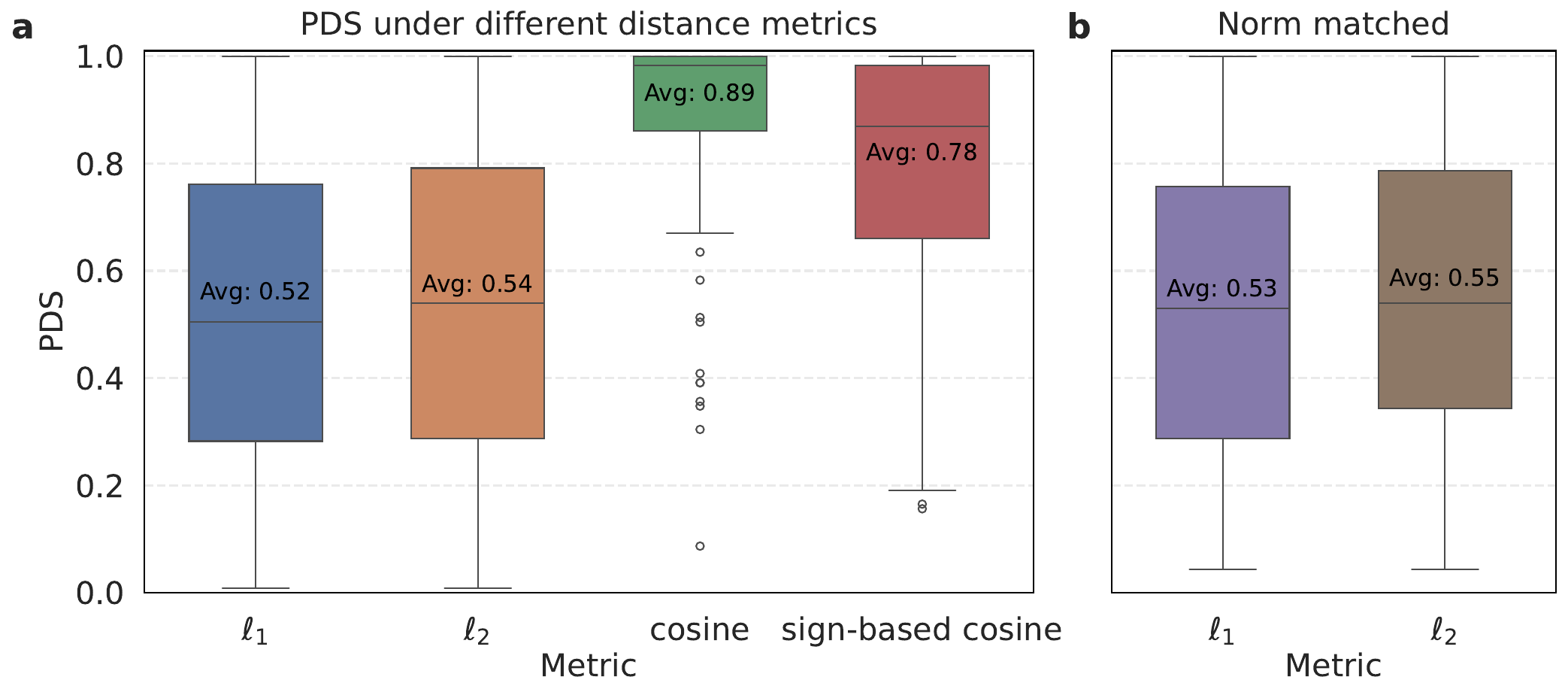}
    \caption{
    \textbf{PDS varies substantially across distance metrics.}
    (a) PDS for the prediction vectors $\hat{\Delta}_i$ computed under four different metrics.
    (b) PDS computed using $\ell_1/\ell_2$ distance after normalizing each prediction to match the corresponding true effect vector in total norm.
In all analyses, the target gene is excluded from each perturbation-effect vector.}
    \label{fig:PDS_distance}
\end{figure}

As shown in Figure~\ref{fig:PDS_distance}a, a notable discrepancy emerges. When PDS is computed using $\ell_1$ or $\ell_2$ distance, the resulting scores are generally low, only marginally above $0.5$, which is close to what one would expect from random guessing. In contrast, when cosine similarity or sign-based cosine similarity is used, the PDS increases substantially, approaching values near $0.8$. 

This contrast raises an immediate and interesting question: how can the very same set of predictions appear almost random under $\ell_1/\ell_2$ distances yet become highly discriminative when evaluated with cosine-based measures?

\section*{Rescaling predicted vectors does not rescue $\ell_1/\ell_2$-based PDS}

One might reasonably suspect that the relatively low PDS values obtained under $\ell_1/\ell_2$ distance arise primarily from mismatched global scales of predicted and observed perturbation effects. To test this possibility, we considered rescaled predictions
\[
\tilde{\Delta}_i = c_i \hat{\Delta}_i,
\]
where $c_i$ is chosen such that $\|\tilde{\Delta}_i\|_{1} = \|\Delta_i\|_{1}$ (for $\ell_1$) or $\|\tilde{\Delta}_i\|_{2} = \|\Delta_i\|_{2}$ (for $\ell_2$). This enforces exact norm matching between each prediction and its true perturbation effect.

However, as shown in Figure~\ref{fig:PDS_distance}b, even after this normalization, the resulting PDS scores under $\ell_1$ and $\ell_2$ remain close to $0.5$ on average. Thus, simply correcting the global scale does not substantially improve discriminability when PDS is defined using magnitude-sensitive metrics.

\begin{figure}[ht]
    \centering
    \includegraphics[width=0.5\linewidth]{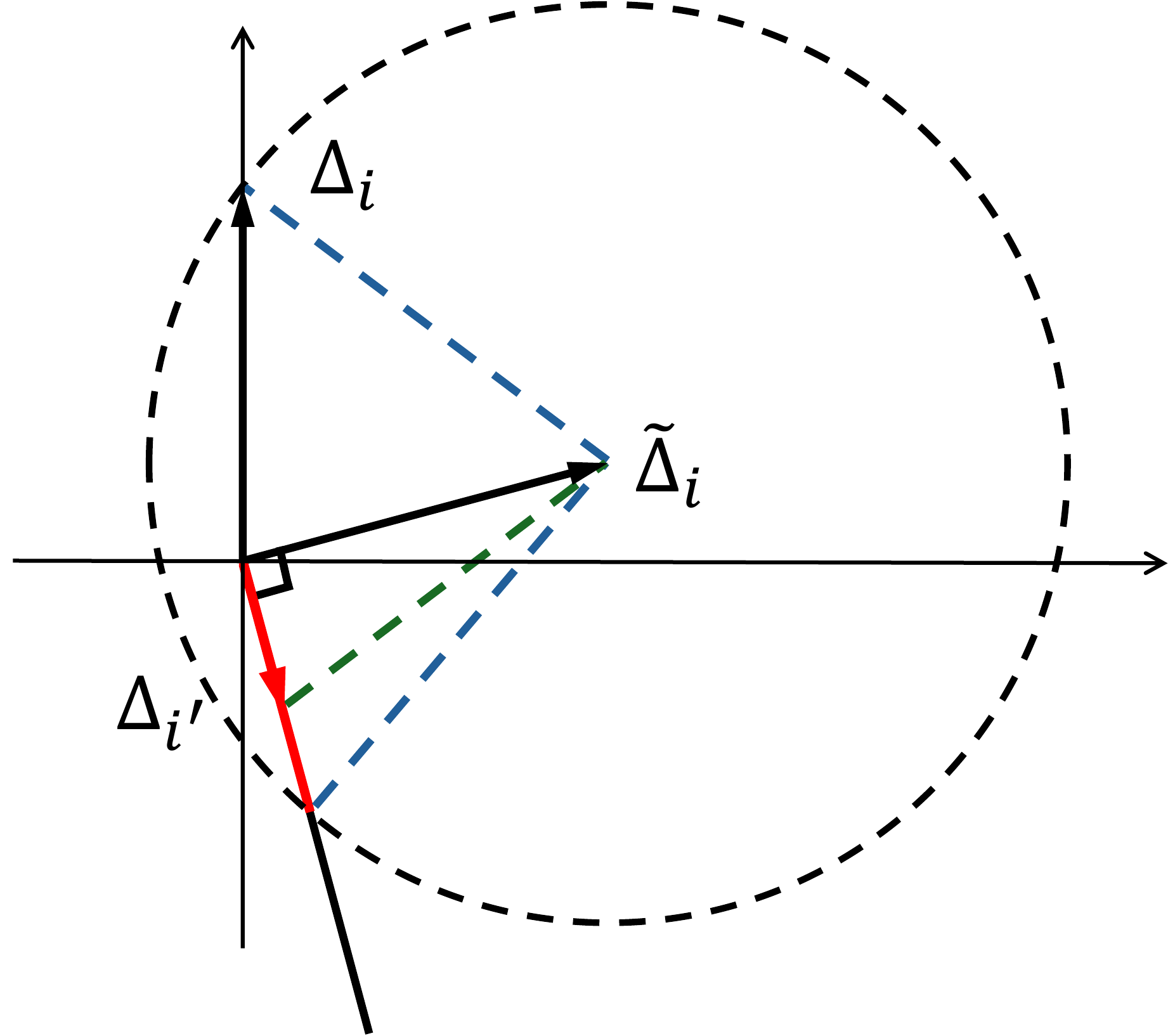}
    \caption{
    \textbf{Geometric illustration of the sensitivity of $\ell_2$-based PDS in two dimensions.}
    Even though the rescaled predicted effect $\tilde{\Delta}_i$ is orthogonal to $\Delta_{i'}$, the shorter magnitude of $\Delta_{i'}$ (red segment) places it closer in Euclidean distance to $\tilde{\Delta}_i$ than $\Delta_{i}$. This demonstrates how $\ell_2$-based rankings can favor vectors with smaller norms despite poorer directional alignment.}
    \label{fig:geometry}
\end{figure}

The underlying reason is geometric. Consider the $\ell_2$ case. Even when the rescaled prediction has the correct length and is more directionally aligned with the true effect $\Delta_i$ than with any other perturbation, the $\ell_2$ distance to another perturbation $\Delta_{i'}$ may still be smaller if $\Delta_{i'}$ has a sufficiently short norm. Figure~\ref{fig:geometry} illustrates this in two dimensions: although $\Delta_{i'}$ is orthogonal to $\tilde \Delta_i$, its shorter length creates a ``red'' segment that lies entirely within the circle of radius $d_{\ell_2}(\tilde{\Delta}_i, \Delta_i)$. Points along this segment remain closer to the prediction in Euclidean distance despite having a larger angular deviation from it.

In this 2D setting, even if $\tilde{\Delta}_i$ is orthogonal to $\Delta_{i'}$, we can guarantee that
\[
d_{\ell_2}(\tilde{\Delta}_i, \Delta_i) < d_{\ell_2}(\tilde{\Delta}_i, \Delta_{i'})
\]
for every $\Delta_{i'}$ (i.e., the circle around $\tilde{\Delta}_i$ does not intersect the ray in the direction of $\Delta_{i'}$) only when
\[
\cos\bigl(\tilde{\Delta}_i, \Delta_i\bigr) > \cos(60^\circ) = 0.5.
\]
In higher dimensions, this effect becomes even more pronounced: short vectors occupy an increasingly large region in which they are closer in $\ell_1/\ell_2$ distance, making magnitude-sensitive versions of PDS relatively insensitive to directional accuracy unless the cosine similarity is very high.

\section*{Why $\ell_1/\ell_2$-based PDS is scale-sensitive and why this behavior is intrinsic}

Another striking feature of $\ell_1/\ell_2$-based PDS is its sensitivity to the overall scale of the predicted vectors. In practice, multiplying all predicted effects $\hat \Delta_i$ by a constant $c$ can substantially change the resulting PDS values, even though scaling leaves all directional information unchanged. This behavior can be explained through simple asymptotic calculations.

\textbf{The $\ell_2$ case.} 
Consider the squared $\ell_2$ distance between a scaled prediction $c \hat{\Delta}_i$ and another perturbation effect $\Delta_{i'}$:
\begin{align*}
d_{\ell_2}\bigl(c\hat{\Delta}_i,\Delta_{i'}\bigr)^2
&= \|c\hat{\Delta}_i - \Delta_{i'}\|_2^2 
= c^2 \|\hat{\Delta}_i\|_2^2 + \|\Delta_{i'}\|_2^2 - 2c \hat{\Delta}_i^\top \Delta_{i'}.
\end{align*}
The difference in squared distances between the true perturbation $i$ and another perturbation $i'$ is therefore
\begin{align*}
& d_{\ell_2}\bigl(c\hat{\Delta}_i,\Delta_{i}\bigr)^2 - d_{\ell_2}\bigl(c\hat{\Delta}_i,\Delta_{i'}\bigr)^2 
= \bigl(\|\Delta_{i}\|_2^2 - \|\Delta_{i'}\|_2^2\bigr) - 2c \bigl(\hat{\Delta}_i^\top \Delta_{i} - \hat{\Delta}_i^\top \Delta_{i'}\bigr).
\end{align*}
Dividing by $c$ and taking $c \to \infty$ yields
\[
\lim_{c \to \infty} \frac{d_{\ell_2}\bigl(c\hat{\Delta}_i,\Delta_{i}\bigr)^2 - d_{\ell_2}\bigl(c\hat{\Delta}_i,\Delta_{i'}\bigr)^2}{c}
= -2 \bigl(\hat{\Delta}_i^\top \Delta_{i} - \hat{\Delta}_i^\top \Delta_{i'}\bigr).
\]
Using the identity
\[
\hat{\Delta}_i^\top \Delta_{i'} = \|\hat{\Delta}_i\|_2 \, \|\Delta_{i'}\|_2 \, \cos\bigl(\hat{\Delta}_i,\Delta_{i'}\bigr),
\]
the condition for the true perturbation $i$ to be ranked closer than $i'$ in the limit $c\to \infty$ becomes
$$\cos\bigl(\hat{\Delta}_i,\Delta_{i}\bigr) > \frac{\|{\Delta}_i'\|_2}{\|{\Delta}_i\|_2}\cos\bigl(\hat{\Delta}_i,\Delta_{i'}\bigr).$$
This inequality shows that the comparison becomes increasingly governed by directional alignment (cosine similarity) as scaling increases. In particular, if $\cos(\hat{\Delta}_i,\Delta_{i'})=0$, meaning the prediction is orthogonal to the wrong perturbation, then the condition reduces to 
$\cos(\hat{\Delta}_i,\Delta_{i})>0$, regardless of the relative magnitudes $\|\Delta_i\|_2$ and $\|\Delta_i'\|_2$. More generally, when $\cos(\hat{\Delta}_i,\Delta_{i'})$ is near $0$, which is common in high-dimensional settings, only modest positive cosine similarity with the true perturbation is needed to outrank many alternatives. Thus, scaling amplifies the directional term relative to magnitude differences, causing the $\ell_2$-based ranking to behave increasingly like cosine similarity. 

\textbf{The $\ell_1$ case.}
A similar limit arises for the $\ell_1$ norm. For each coordinate $j$,
\[
\bigl|c\hat{\Delta}_{ij} - \Delta_{i'j}\bigr|
= c \bigl|\hat{\Delta}_{ij}\bigr| - \operatorname{sign}(\hat{\Delta}_{ij}) \operatorname{sign}({\Delta}_{i'j})\bigr|\Delta_{i'j}\bigr|
\quad (c \to \infty),
\]
so that
\begin{align*}
& \lim_{c\to \infty}\left\{d_{\ell_1}\bigl(c\hat{\Delta}_i,\Delta_{i}\bigr) - d_{\ell_1}\bigl(c\hat{\Delta}_i,\Delta_{i'}\bigr)\right\}
= -\Bigr[\sum_{j=1}^p \operatorname{sign}(\hat{\Delta}_{ij})\operatorname{sign}({\Delta}_{ij}) \bigr|\Delta_{ij}\bigr| - \sum_{j=1}^p \operatorname{sign}(\hat{\Delta}_{ij})\operatorname{sign}({\Delta}_{i'j}) \bigr|\Delta_{i'j}\bigr| \Bigl].
\end{align*}
 Thus, in this limit, $\ell_1$-based PDS becomes driven by a weighted sign cosine similarity. This links $\ell_1$-based PDS to a form of sign-based similarity, in which agreement in sign on large-magnitude coordinates contributes the most to the ranking. 

\begin{figure}[ht]
    \centering
    \includegraphics[width=0.8\linewidth]{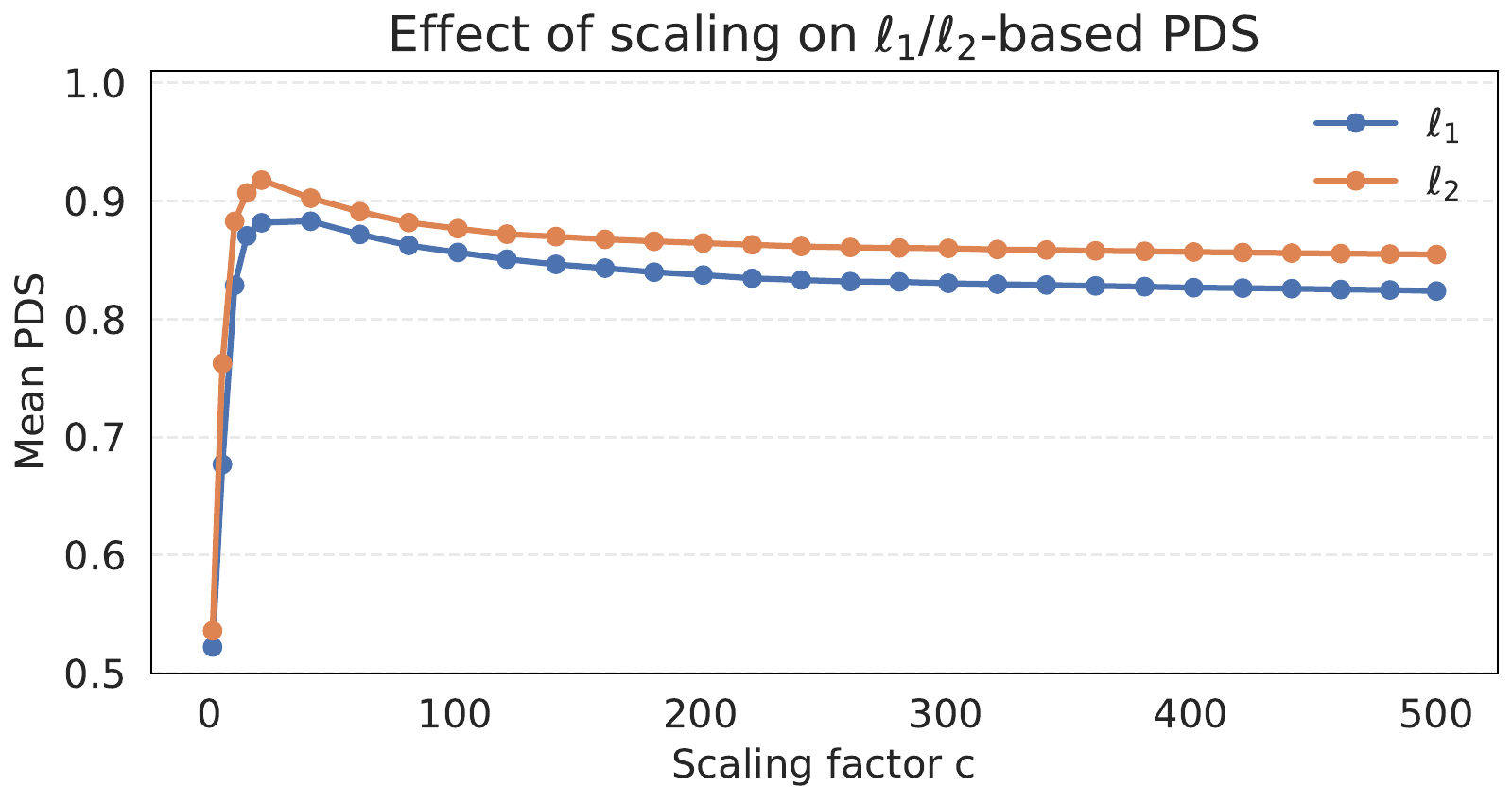}
    \caption{
    \textbf{$\ell_1/\ell_2$-based PDS metrics are sensitive to the magnitude of predicted effects.}
    Mean PDS of scaled predictions $c\hat{\Delta}_{i}$ vary with scaling factor $c$.
    The blue and orange curves correspond to PDS computed with $\ell_1$ and $\ell_2$ distance, respectively.
    }
    \label{fig:PDS_scale}
\end{figure}

These asymptotic results show that the scale dependence of $\ell_1/\ell_2$-based PDS is mathematically predictable. As predictions are globally rescaled, PDS rankings transition toward those determined by cosine or sign-based cosine similarities. Figure~\ref{fig:PDS_scale} illustrates this process: as the scaling factor $c$ increases, $\ell_1/\ell_2$-based PDS values rise sharply and eventually plateau at a limiting value. 

Importantly, this is not an artifact of ``hacking'' the scoring metric. Instead, it is an inherent property of magnitude-sensitive norms in multi-dimensional spaces, where distances combine both scale and direction. Scaling modifies the magnitude component while leaving direction unchanged, causing $\ell_1/\ell_2$-based PDS to interpolate toward cosine or sign-based behavior.

\section*{Implications for future PDS-style metrics}

The analyses above suggest two potential directions for refining PDS in future perturbation-effect benchmarks.

A first option is to define PDS using cosine-based similarity measures. Because cosine similarity depends only on direction, these metrics are invariant to global rescaling and robust to differences in normalization or preprocessing. Cosine-based PDS directly assesses whether predicted perturbation effects capture the correct pattern of up- and down-regulated genes, without being influenced by inconsistencies in total effect magnitude across studies or platforms.

A natural concern is that cosine-based PDS may be easier to score well on: achieving high discrimination does not necessarily require high correlation with the true effects, only that predictions outperform random guessing. A stricter alternative is therefore to retain $\ell_1/\ell_2$-based PDS,  but to fix the norm of the predicted vectors -- for example by enforcing $\|\hat{\Delta}_i\|_1 = \|\Delta_i\|_1$ or $\|\hat{\Delta}_i\|_2 = \|\Delta_i\|_2$ before computing distances. This adjustment does not turn PDS into a measure of magnitude accuracy; instead, it removes the possibility of improving scores through arbitrary rescaling. Under this ``norm-matched'' PDS, achieving high discrimination requires genuinely close directional alignment with the true perturbation effects, yielding a more stringent and interpretable metric.

\section*{Why PDS should focus on directional accuracy rather than total magnitude}

Neither of the refinements discussed above requires PDS to evaluate the total magnitude accuracy of the predicted perturbation effects, and in many settings, this is a desirable property. Total effect magnitudes are intrinsically difficult to predict: they depend on guide RNA efficiency, experimental design, noise levels in single-cell RNA-seq, and even subtle preprocessing choices.

\begin{figure}[ht]
    \centering
    \includegraphics[width=\linewidth]{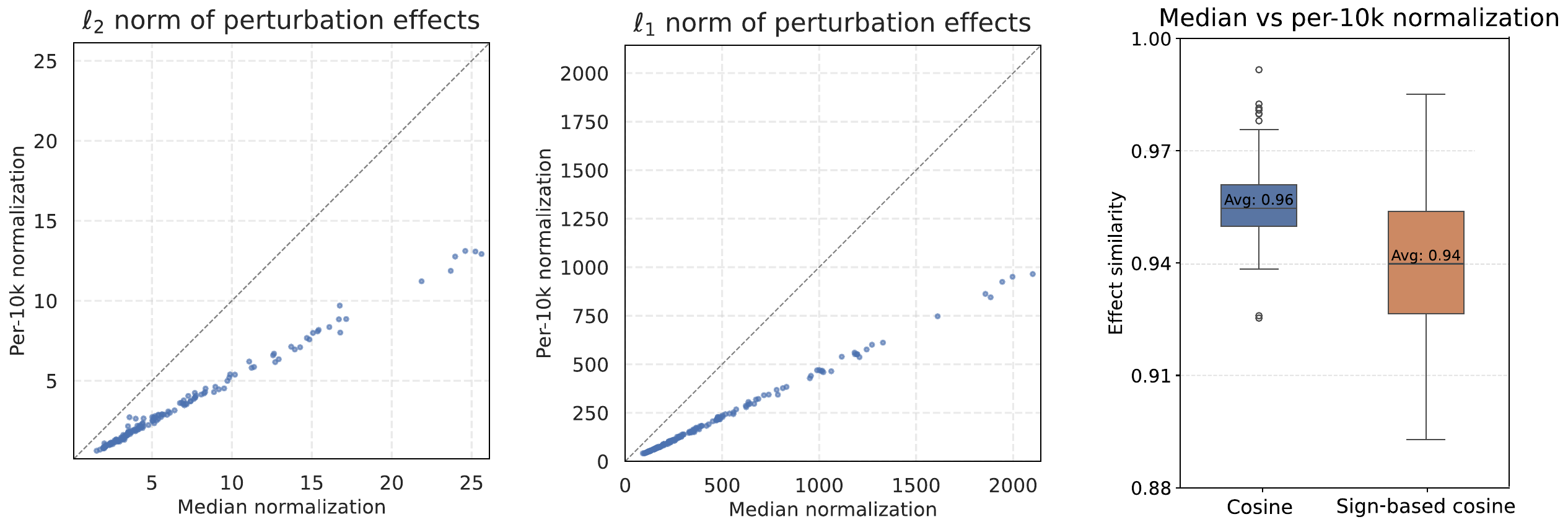}
    \caption{
\textbf{Perturbation-effect magnitudes depend strongly on preprocessing.}
    The first two panels show the $\ell_2$ and $\ell_1$ norms of mean perturbation effects obtained using per-10k versus median normalization; each point corresponds to one perturbation.
The third panel shows boxplots of the cosine and sign-based cosine similarities between the effects produced by the two preprocessing pipelines.
    }
    \label{fig:preprocessing}
\end{figure}

In practice, the total $\ell_1$ and $\ell_2$ norms of perturbation-effect vectors can vary dramatically under different normalization pipelines. For example, Figure~\ref{fig:preprocessing} compares observed mean perturbation effects derived from the same raw counts of the VCC training data after two common preprocessing choices: (i) median library-size normalization and (ii) per-10k scaling followed by log1p transformation. Although these transformations yield perturbation effects with high cosine similarity, their total $\ell_1$ and $\ell_2$ norms differ substantially.%

These observations suggest that direction-based metrics may provide more stable and meaningful comparisons across contexts, particularly when integrating data across experiments or platforms. In contrast, $\ell_1/\ell_2$-based PDS is sensitive to global scaling in a way that does not consistently reflect biological magnitude accuracy: increasing the scale of the predicted effects can artificially inflate the score, while incorrect magnitudes may be penalized or rewarded depending on their interaction with vector norms and angular relationships. Thus, even if one wished to evaluate total magnitude accuracy, $\ell_1/\ell_2$-based PDS would not be the appropriate tool. Focusing PDS on directional information avoids this instability and more directly evaluates the aspect of perturbation effects that is most reproducible across contexts.

\section*{Conclusion}
Our analysis clarifies how the Perturbation Discrimination Score behaves under different similarity measures and under global scaling of predictions. In particular, $\ell_1/\ell_2$-based PDS can be strongly influenced by the magnitude of predicted vectors, while cosine-based versions are more robust to these effects. These findings do not critique the use of PDS in benchmarking the prediction performance; rather, they illuminate the metric’s geometry and offer guidance for designing future discriminability-based evaluations. As perturbation-prediction benchmarks continue to grow in scale and complexity, a deeper understanding of PDS-style metrics will help ensure that evaluation criteria reflect the biological and statistical goals of each task.

\vspace{1em}

\bibliography{ref} 
\bibliographystyle{apalike}

\end{document}